\documentclass[12pt]{article}
\usepackage[utf8]{inputenc}
\usepackage{amsmath,amsfonts,amsthm}
\usepackage{amssymb}
\usepackage{geometry}
\usepackage[superscript]{cite}
\usepackage{graphicx}
\usepackage{caption}
\usepackage{subcaption}
\usepackage{setspace}
\usepackage{lipsum}
\usepackage{mathtools}
\usepackage{cuted}
\usepackage[english]{babel}
\usepackage{hyperref}
\usepackage{multicol}
\usepackage[dvipsnames]{xcolor}
\usepackage{multirow}
\usepackage{multirow}
\usepackage{tabularx}
\usepackage{booktabs}
\usepackage{multicol}
\usepackage{mhchem}
\usepackage{rotating}
\usepackage{booktabs}
\usepackage{array}
\usepackage{makecell}
\usepackage{multirow}
\usepackage{pifont}
\usepackage{amssymb}
\usepackage{fontawesome}

\title{\textbf{Multigeometric Breathing Mode Framework for Viruses}}
\author{Krish Bardhan$^{1}$, Charu Sharma$^{1}$, Shivansh Bhatnagar$^{1}$, and Prafulla K Jha$^{1,*}$}
\date{\small$^{1}$\textit{Department of Physics, Faculty of Science, The Maharaja Sayajirao University of Baroda, Vadodara, Gujarat 390002, India.}\\ \footnotesize Corresponding Authors: $^{*}$prafullaj@yahoo.com}

\begin{document}

\maketitle
\doublespacing

\begin{abstract}
	
\doublespacing

 The estimation of the breathing mode frequency for viral capsids remains a persistent challenge across literature which spawned various frameworks and methodologies. However, discrepancies were observed between the experimental Low Frequency Raman Scattering (LFRS) and calculated pre-existing values. To solve this discrepancy, we propose a framework which is developed to determine the breathing-mode frequency by modelling the virus as a macroscopic coupled harmonic oscillator. By integrating mass-loading directly into the classical elastodynamic equations, this model yields an analytical expression that couples the system's total inertia with its specific geometry. Moreover, the viruses are segregated according to their geometries into three coordinates to acquire their respective geometric eigen values. Here we illustrate how the proposed Multigeometric Breathing mode Framework for Virus (MBFV) outperforms prior models by yielding closer values than the pre-existing frameworks upon validating against LFRS.
 
\end{abstract}
\
\textit{keywords} : Viral capsids, Breathing mode,  Low-Frequency Raman Spectra(LFRS), Acoustic Radiation

\section{Introduction}
Viruses are mechanically active anomalies that profoundly turn inertial cells into viral replication centers by sabotaging their regular functioning. The viral capsid accomplishes this task by lurking in the nucleic acid of the host cell to alter its structure to obtain a sustainable environment to inexorably infect host cells  [1-5].

Owing to their mechanical activity, they exhibit the tendency to store elastic stress, support collective vibrational motion, and undergo conformational fluctuations relevant to infectivity. These features make the viral capsid both an object of physical study and a target of diagnostic and therapeutic interest[1,2,5,6]. However, this renders difficulty in detecting and eliminating the viral structure with conventional approaches because its mechanically relevant states are collective, environment-dependent, and often hidden beneath large thermal and solvent backgrounds[3-7]. This challenge is further amplified by the fact that the viral capsids are typically heterogeneous, faceted, hydrated, and under prestress, rather than ideal homogeneous solids[3,5,6]. Subsequently, their response to deformation depends on shell thickness, internal genome loading, electrostatic effects, solvent coupling, and icosahedral symmetry breaking, all of which can shift the characteristic vibrational spectrum away from the simplest continuum predictions[3-18]. As a result, identifying a virus mechanically is not merely a matter of measuring size or mass; it requires resolving the collective eigenmodes of the capsid and understanding how those modes change with geometry and environment[7,9]. Among these eigenmodes, the most fundamental is the spheroidal breathing mode ($l=0$), a purely radial oscillation in which the capsid expands and contracts coherently as a whole. Since this mode represents the lowest-order volumetric deformation, it is the closest mechanical analogue of a global pulse of the viruses, and it is therefore central to both resonance-based detection and destabilization respectively[9-11,19-25].

The breathing mode is best understood in the language of eigen-frequencies and resonant frequency. In linear elasto-dynamics, a structure has natural frequencies determined by its mass distribution, geometry, and elastic constants; when driven near one of these frequencies, the response amplitude rises sharply and is only limited by damping[3,9,23-26]. Moreover, for a viral capsid, the breathing-mode frequency is the lowest spheroidal eigen-frequency (usually in the GHz range), and is precisely the mode most strongly associated with coherent radial compression and expansion[10,11,13,25-33]; making low-frequency Raman spectroscopy and related ultrafast optical methods very useful as they couple to Raman-active collective modes which can reveal the signature of the [$l=0$] vibration as a measurable spectral feature[9-17,24-30,32-42]. Accordingly, the breathing-mode frequency functions as a mechanical fingerprint, one that can complement biochemical assays and be especially valuable when the goal is intrinsic detection of intact viral particles[4-13,24-29]. From an elimination perspective, driving a virus near its breathing mode resonance ($l=0$) is vital, as it channels energy into a uniform global deformation, thereby increasing the likelihood of mechanical instability throughout the capsid [10,25–28,43]. Capsid mechanics demonstrate that perturbing these vibrational states beyond stability limits causes viral rupture and loss of infectivity [1,3,8,10–14,44]. However, therapeutic resonance demands caution; selectivity relies heavily on precise frequency targeting, acoustic coupling, the local environment, and field intensity [1,3,8,10–19].  Consequently, the breathing mode constitutes a preferred target because its role as the dominant global mode renders it easier to identify and model than the higher order quadrupolar modes, which exhibit greater sensitivity to asymmetry, solvent perturbations, and structural disorder[9,10,14].

To model the breathing mode of viruses, many attempts have been made. These attempts(frameworks) can generally be divided into three generations. The first generation frameworks mainly consist of continuum elastic models that ardently abide by Lamb's equations [3,10,11,24,27], yielding spheroidal modes through the Elastodynamic Navier-Cauchy equation [1,3,10,12,23], where frequency scales inversely with radius and longitudinal sound speed [1,3,45]. However, real viruses depart from this idealization [5-8,43,46], making the extracted frequency an effective collective frequency [6,7,9,31] and driving the literature to balance analytic tractability against biological realism [7-9,13]. Because models based on Lamb's equations inherently neglect the shell's morphology and aqueous medium, they lead to a systemic overestimation of the breathing mode frequency [47-53]. Free shell extensions improve on this by introducing fluid boundaries and complex frequencies, yet they remain too idealized for predictive use by treating the capsid crudely as a homogeneous solid sphere. Similarly, elastic network models replace continuum elasticity with a spring network built from protein coordinates [1-7,24-31, 47-49] to identify which parts move together, but they rely on an ad hoc universal spring constant and cannot naturally produce damping or solvent effects.  The second-generation family addresses these deficiencies by retaining much more structural detail through structural awareness methods such as all-atom molecular dynamics (MD), coarse-grained MD, and finite-element frameworks [4,7,14,17,20,21,26,54]. While all-atom MD is highly physical, its substantial computational cost makes it rarely practical for routine comparative studies. Coarse-grained MD reduces this burden by mapping many atoms into fewer interaction sites to make longer timescales accessible [32,55], but trade-offs include a loss of atomic detail, parameter ambiguity, and degraded fidelity that can yield errors scaling from 20\% to 50\%. Consequently, finite-element analysis sits between continuum theory and detailed structure by incorporating explicit meshing [26], yet this method remains sensitive to assumed constants and boundary conditions, and is often hampered by severe computational bottlenecks that make routine spectral analysis practically infeasible.  Subsequently, third-generation approaches encompass recent data-driven hybrid and multiscale models such as normal-mode analysis, block-based reduced models, and layered or core-shell extensions—attempting to compress the high-dimensional problem into a reduced but physically meaningful basis [9,13,43,47,48,55-59]. Normal-mode analysis is effective for low-frequency collective motions, but its harmonic approximation suppresses anharmonicity and provides no intrinsic damping, failing to describe experimentally broadened resonances [24-30, 36-42]. Block-based reduced models improve efficiency by projecting dynamics onto selected capsid subspaces [58], but they remain valid only below the intrinsic high-frequency limit of capsomeres, beyond which atomic-scale fluctuations necessitate computationally intensive all-atom Hessian-based analyses. To reduce complexity, these models ultimately assume highly symmetric viral architectures, allowing elasticity calculations to be extrapolated across homologous structural elements [60-62]
While layered sphere models alleviate homogeneous-shell shortcomings, they fail to capture discrete molecular architectures, anisotropic mechanics, and hydrodynamic damping [61], ultimately relying on poorly constrained smoothing functions and experimentally inaccessible parameters [34,55,63-72].
Across all three generations, common structural limitations persist, wherein each framework captures one significant aspect of the viral breathing problem while omitting another. Hence, upon examining these ubiquitous limitations, the emergence of a newer model to eliminate these shortcomings became evident.
In this paper, we develop a Multigeometric Breathing Mode Framework for Viruses (MBFV), which incorporates a frequency-dependent viscoelastic correction factor that captures acoustic vibrational damping and solvent coupling arising from the surrounding biological medium. These effects were systematically neglected in previous frameworks.

\section{Methodology}
{\small
	Before delving into the mathematical derivations, it is an absolute necessity to address the reasonings which serve as the inherent underlying foundations with regard to the functioning of MBFV. The mathematical scaffolding of the proposed MBFV originates from the Elastodynamic Navier-Cauchy equation, which was further moulded by Lamb. Albeit, the initiation point of the proposed framework is the aforementioned differential equation is in turn substituted by incorporating and utilising the index notations to simplify and truncate the equation. Viruses are conceptualized as elastic continua whose mechanical behaviours are dictated by the Elastodynamic Navier-Cauchy (Lame) equation[3,9,10],
	
	\begin{equation}
		\rho_s \frac{\partial^2 u_i}{\partial t^2} = (\lambda + \mu) \partial_i (\partial_j u_j) + \mu \partial_j \partial_j u_i
	\end{equation}
	
	Here, $\rho_s \frac{\partial^2 u_i}{\partial t^2}$ represents the inertial force per unit volume, with capsid density $\rho_s$ and displacement vector $u_i$. The right-hand side decomposes elastic restoring forces into two distinct phenomena. The first term, $(\lambda + \mu) \partial_i (\partial_j u_j)$, utilizes Lamé parameters $\lambda$ and $\mu$ to define the gradient of the displacement divergence, driving longitudinal, compressional waves (volumetric dilatation). The second term, $\mu \partial_j \partial_j u_i$, is the displacement Laplacian, governing isochoric, transverse shear waves.
	
	To bypass atomic-scale computational limits, a macroscopic continuum reduction amalgamates discrete masses into a continuous density field $\rho_s$ and Cauchy stress tensor. This Navier Cauchy equation encapsulates the intrinsic elastic restoring forces thereby establishing a framework for modeling biological energy dissipation. While classical Lamb theory artificially isolates the virion via traction-free boundaries, our methodology restructures boundary mechanics to account for the massive inertial drag of the surrounding fluid, ensuring physically accurate fluid-structure coupling.
	
	To Analyze the fundamental radial breathing mode ($l=0$) isolation of the purely radial expansion is required because this mode is a strictly dilatational, and symmetric volumetric oscillations devoid of rotational shearing.[3,9,10]
	The full three-dimensional tensor field is mathematically redundant. To decouple these motions, we apply the Helmholtz Decomposition theorem, resolving the dynamic structural displacement field $u_i$ into an irrotational scalar potential $\psi$ and a solenoidal vector potential $\Psi_k$.
	$\Psi_k$:
	\begin{equation}
		u_i = \partial_i \psi + \epsilon_{ijk} \partial_j \Psi_k
	\end{equation}
	where $\epsilon_{ijk}$ denotes the Levi-Civita permutation symbol. The scalar gradient ($\partial_i \psi$) governs pure volumetric dilatations, while the vector curl ($\epsilon_{ijk} \partial_j \Psi_k$) governs isochoric shear distortions. Since the low-frequency Raman scattering observables strictly correspond to the radial breathing mode, transverse shear components must identically vanish. By setting the rotational vector potential to zero ($\Psi_k=0$), the three-dimensional displacement field collapses into the gradient of a single scalar potential:
	\begin{equation}
		u_i = \partial_i \psi
	\end{equation}
	
	Substituting this into the Navier-Cauchy equation and applying the divergence operator to both sides converts the tensor equation into a scalar form, proving the curl of the gradient to be zero. Eventually, the tensor equation collapses. This mathematically strips away structural shear mechanics, reducing the system to a fundamental scalar wave equation:
	\begin{equation}
		\nabla^2 \psi - \frac{1}{V_I^2} \frac{\partial^2 \psi}{\partial t^2} = 0
	\end{equation}
	Here, $\nabla^2$ signifies the spatial Laplacian operator, and $V_I$ represents the intrinsic longitudinal acoustic wave velocity propagating through the protein medium. This velocity emerges directly from the capsid's uniform mass density and its macroscopic Lamé elastic parameters:
	
	\begin{equation}
		V_I = \sqrt{\frac{\lambda + 2\mu}{\rho_s}}
	\end{equation}
	
	The elimination of rotational dynamics established above confirms that the radial breathing mode propagates as a longitudinal acoustic wave confined within the geometry of the virus carrying no shear, no torsion, and no transverse displacement. This result physically signifies that a purely longitudinal wave in a bounded elastic medium is governed entirely by a single scalar field, whose spatial structure encodes the virus's geometry and whose temporal evolution encodes its resonant frequency. Therefore, the reduction of the full three-dimensional Navier-Cauchy tensor equation to this scalar wave equation is not merely a mathematical convenience, it is the necessary prerequisite that makes the eigen-frequency problem separable, allowing the temporal and spatial degrees of freedom to be decoupled and the characteristic vibrational frequency to be extracted as a direct geometric fingerprint of the capsid.
	
	A virus driven into its fundamental breathing mode vibrates in a distinct normal mode, requiring synchronous motion where every spatial coordinate reaches its maximum volumetric amplitude simultaneously. This physical coherence permits the separation of variables, by decoupling temporal evolution from the static structural configuration. The scalar potential $\phi$ is divided into a spatial amplitude function $\Psi$ and a harmonic temporal function:
	
	\begin{equation}
		\phi(\mathbf{r},t) = \Psi(\mathbf{r})e^{-i\omega t}
	\end{equation}
	
	Since, the temporal evolution is purely oscillatory, the substitution of the harmonic postulation into the scalar wave equation and evaluation of the second-order temporal derivative eliminates the time-dependent factor, reducing the system to a static spatial Helmholtz differential equation: 
	
	\begin{equation}
		\nabla^2\Psi + k^2\Psi = 0
	\end{equation}
	
	Here, the wavenumber $k$ relates the angular frequency $\omega$ to the intrinsic longitudinal acoustic wave velocity $V_L$:
	
	\begin{equation}
		k = \frac{\omega}{V_L}
	\end{equation}
	
	This Helmholtz equation governs the volumetric displacement amplitude across the three-dimensional structure of the virus. As much as these structures vary, exhibiting spherical, filamentous, or ellipsoidal conformations, the expansion of the spatial Laplacian $\nabla^2$ into appropriate coordinates, assuming a fully separable spatial wavefunction $\Psi$:
	
	\begin{equation}
		\Psi(r,\theta,z) = R(r)\Theta(\theta)Z(z)
	\end{equation}
	
	For the primary radial breathing mode, the azimuthal angular dependence vanishes identically ($m=0$). This yields a Bessel differential equation for filamentous virions,  governed by the zeroth-order Bessel function of the first kind, $J_0(\alpha r)$. For icosahedral structures in the spherical limit, spatial mechanics are governed by the spherical Bessel function $j_0(kr)$.
	
	Extracting the fundamental roots of these geometric functions requires the invocation of temporary traction-free boundary conditions at the outermost viral radius $R$:
	
	\begin{equation}
		\sigma_{rr}(R) = 0
	\end{equation}
	
	These idealized constraints yield fundamental geometric roots of $\xi_0 \approx 5.763$ for spherical coordinates and $\alpha R \approx 2.405$ for cylindrical coordinates. Since restricting the framework to highly symmetric geometries introduces errors for irregular viruses, hence a Cartesian coordinate system assumes a separable spatial wavefunction $\Phi(x,y,z)$:
	
	\begin{equation}
		\Phi(x,y,z) =
		\sin(k_x x)\sin(k_y y)\sin(k_z z)
	\end{equation}
	
	The fundamental eigenvalue resolves into the sum of squared wavenumbers:
	
	\begin{equation}
		\Lambda_{\mathrm{geom}} =
		k_x^2 + k_y^2 + k_z^2
	\end{equation}
	
	The singular parameter $\Lambda_{\mathrm{geom}}$ encapsulates this geometric convergence and adapts by embedding  target viruses into the appropriate coordinates:
	
	\begin{equation}
		\begin{aligned}
			\Lambda_{\mathrm{geom}}^{(\mathrm{spherical})}
			&= \left(\frac{5.76}{R}\right)^2 \\
			\Lambda_{\mathrm{geom}}^{(\mathrm{cylindrical})}
			&= \left(\frac{2.405}{R}\right)^2 \\
			\Lambda_{\mathrm{geom}}^{(\mathrm{cartesian})}
			&= k_x^2 + k_y^2 + k_z^2
		\end{aligned}
	\end{equation}
	
	To actualize this spatial fingerprint in a biological environment, the virion is treated as a fully coupled fluid-structure harmonic oscillator. This establishes the energetic foundation and mathematical scaffolding necessary to introduce dissipative energy losses, modeling the damping phenomena observed in viscoelastic biological media. The breathing mode is regulated by the total mechanical energy $E$, requiring a balance between kinetic energy $T$ and potential energy $U$:
	
	\begin{equation}
		E = T + U
	\end{equation}
	The potential energy $U$ stored within the protein capsid, determined by the internal elastic strain, is defined via the effective elastic stiffness $K_{\mathrm{eff}}$ and the macroscopic radial displacement $R$:
	
	\begin{equation}
		U = \frac{1}{2}K_{\mathrm{eff}}R^2
	\end{equation}
	
	Conversely, the kinetic energy $T$, representing the total inertial resistance during volumetric oscillation, is regulated by the total effective mass $M_{\mathrm{eff}}$ and the temporal derivative of radial displacement $\dot{R}$:
	
	\begin{equation}
		T = \frac{1}{2}M_{\mathrm{eff}}\dot{R}^2
	\end{equation}
	
	The total effective mass $M_{\mathrm{eff}}$ incorporates the mass of the elastic capsid $M_s$, the internal genomic core $M_c$, and the hydrodynamic added mass of the external medium $M_f$:
	
	\begin{equation}
		M_{\mathrm{eff}} = M_s + M_c + M_f
	\end{equation}
	
	Equating these energies yields the squared angular frequency $\omega^2$. Normalizing the expression by the isolated capsid mass $M_s$ mathematically demonstrates how structural mass loading lowers the frequency due to inertia:
	
	\begin{equation}
		\omega^2 = \frac{K_{\mathrm{eff}}}{M_{\mathrm{eff}}}
	\end{equation}
	
	\begin{equation}
		\omega^2 =
		\frac{K_{\mathrm{eff}}}{M_s}
		\frac{1}{1 + \frac{M_f}{M_s} + \frac{M_c}{M_s}}
	\end{equation}
	
	Formalizing these parameters across the continuum requires explicit volumetric integration. Substituting harmonic temporal velocity derivatives yields the maximum kinetic energy:
	
	\begin{equation}
		T_{\mathrm{max}} =
		\frac{1}{2}(2\pi\omega)^2
		\left(
		\int_{V_c} \rho_c \lvert u_c \rvert^2\,dV
		+
		\int_{V_s} \rho_s \lvert u_s \rvert^2\,dV
		+
		\int_{V_f} \rho_f \lvert u_f \rvert^2\,dV
		\right)
	\end{equation}
	
	Concurrently, the maximum potential energy is governed entirely by the restoring forces of the protein shell, adhering to standard linear elasticity principles:
	
	\begin{equation}
		U_{\mathrm{max}} =
		\frac{1}{2}
		\int_{V_s}
		\sigma_{ij}\epsilon_{ij}\,dV
	\end{equation}
	
	Applying Rayleigh's energetic quotient by equating $T_{\mathrm{max}}$ and $U_{\mathrm{max}}$ yields the squared frequency:
	
	\begin{equation}
		\omega^2 =
		\frac{1}{(2\pi)^2}
		\frac{
			\displaystyle \int_{V_s}\sigma_{ij}\epsilon_{ij}\,dV
		}{
			\displaystyle
			\int_{V_s}\rho_s\lvert u_s\rvert^2\,dV
			+
			\int_{V_c}\rho_c\lvert u_c\rvert^2\,dV
			+
			\int_{V_f}\rho_f\lvert u_f\rvert^2\,dV
		}
	\end{equation}
	
	Dividing both the numerator and denominator by the isolated kinetic energy integral of the unperturbed shell,
	$\int_{V_s}\rho_s\lvert u_s\rvert^2\,dV$,
	transforms the numerator into the squared classical frequency. Replacing this term with the spatial eigenvalue $\Lambda_{\mathrm{geom}}$ and longitudinal acoustic velocity $V_I$ yields the unperturbed classical limit:
	
	\begin{equation}
		\left(\frac{V_I}{2\pi}\right)^2
		\omega_{\mathrm{vac}}^2 =
		\Lambda_{\mathrm{geom}}
	\end{equation}
	
	The denominator concurrently becomes a dimensionless summation representing relative mass loading, yielding the unified frequency equation:
	
	\begin{equation}
		\omega =
		\frac{V_I}{2\pi}
		\sqrt{
			\frac{\Lambda_{\mathrm{geom}}}{
				1
				+
				\frac{
					\displaystyle \int_{V_c}\rho_c\lvert u_c\rvert^2\,dV
				}{
					\displaystyle \int_{V_s}\rho_s\lvert u_s\rvert^2\,dV
				}
				+
				\frac{
					\displaystyle \int_{V_f}\rho_f\lvert u_f\rvert^2\,dV
				}{
					\displaystyle \int_{V_s}\rho_s\lvert u_s\rvert^2\,dV
				}
			}
		}
	\end{equation}
	
	Evaluating these specific mass distributions require volumetric integration of their respective densities. The intrinsic mass of the unperturbed capsid shell, defined by its uniform density $\rho_s$ and outer radius $R_o$, is:
	
	\begin{equation}
		M_s = \rho_s\frac{4\pi R_o^3}{3}
	\end{equation}
	
	Since the encapsulated genetic material couples mechanically to the capsid's internal surface, rather than undergoing uniform volumetric deformation, its mass ratio scales by the squared ratio of the inner radius $R_i$ to outer radius $R_o$:
	
	\begin{equation}
		\frac{M_c}{M_s}
		=
		\frac{\rho_c}{\rho_s}
		\left(\frac{R_i}{R_o}\right)^2
	\end{equation}
	
	For oscillatory radial potential flow, the external medium acts as a classical hydrodynamic added mass. Evaluating this velocity field simplifies the fluid mass $M_f$ and its ratio to the shell:
	
	\begin{equation}
		M_f = \rho_f\frac{4\pi R_o^3}{3}
	\end{equation}
	
	\begin{equation}
		\frac{M_f}{M_s}
		=
		\frac{\rho_f}{\rho_s}
	\end{equation}
	
	To accommodate dissipation within a compressible, viscoelastic environment, the structural wave equation expands to include the elasticity tensor $c_{ijkl}$, effective density $\rho_{\mathrm{eff}}(\omega)$, and a frequency-dependent memory kernel $\Gamma(\omega)$ illustrating viscous damping:
	
	\begin{equation}
		\nabla \cdot
		\left(
		c_{ijkl}\partial_k\partial_l\Psi
		\right)
		+
		\omega^2\rho_{\mathrm{eff}}(\omega)\Psi
		+
		i\omega\Gamma(\omega)\Psi
		=0
	\end{equation}
	
	The memory kernel defines damping via shear viscosity $\eta$ and viscoelastic relaxation time $\tau$:
	
	\begin{equation}
		\Gamma(\omega)
		=
		\frac{4\eta\rho_s}{3}
		\frac{\omega}{1+i\omega\tau}
	\end{equation}
	
	Hydrodynamic loading thus transitions into a frequency-dependent parameter $\chi(\omega)$, accounting for acoustic radiation and viscous dissipation via the fluid sound speed $c_f$ and relaxation constant $\delta$:
	
	\begin{equation}
		\chi(\omega)
		=
		\frac{\rho_f}{\rho_s}
		\left[
		1
		+
		\frac{i\omega R_o}{c_f}
		+
		\frac{4\eta}{3\rho_f c_f^2}
		\frac{i\omega}{1+i\omega\delta}
		\right]
	\end{equation}
	
	For a continuum approximation prioritizing purely inertial mass-loading without extreme viscoelastic dispersion, this correction factor reduces to its baseline scalar form $\chi$:
	
	\begin{equation}
		\chi =
		\frac{\rho_c}{\rho_s}
		\left(\frac{R_i}{R_o}\right)^2
		+
		\frac{\rho_f}{\rho_s}
	\end{equation}
	
	Using this correction factor, the specific breathing mode frequencies for target coordinate geometries are evaluated as:
	
	\textbf{Spherical Coordinates:}
	
	\begin{equation}
		\omega_{\mathrm{sph}}
		=
		\frac{V_l}{2\pi}
		\sqrt{
			\frac{
				\left(\dfrac{5.76}{R}\right)^2
			}{
				1
				+
				\dfrac{\rho_c}{\rho_s}
				\left(\dfrac{R_i}{R_o}\right)^2
				+
				\dfrac{\rho_f}{\rho_s}
			}
		}
	\end{equation}
	
	\textbf{Cylindrical Coordinates:}
	
	\begin{equation}
		\omega_{\mathrm{cyl}}
		=
		\frac{V_l}{2\pi}
		\sqrt{
			\frac{
				\left(\dfrac{2.405}{R}\right)^2
			}{
				1
				+
				\dfrac{\rho_c}{\rho_s}
				\left(\dfrac{R_i}{R_o}\right)^2
				+
				\dfrac{\rho_f}{\rho_s}
			}
		}
	\end{equation}
	
	\textbf{Cartesian Coordinates:}
	
	\begin{equation}
		\omega_{\mathrm{cart}}
		=
		\frac{V_l}{2\pi}
		\sqrt{
			\frac{
				k_x^2+k_y^2+k_z^2
			}{
				1
				+
				\dfrac{\rho_c}{\rho_s}
				\left(\dfrac{R_i}{R_o}\right)^2
				+
				\dfrac{\rho_f}{\rho_s}
			}
		}
	\end{equation}
	
	Substituting the comprehensive frequency-dependent correction factor yields the generalized unified equation for virion breathing modes:
	
	\begin{equation}
		\omega =
		\frac{V_I}{2\pi}
		\sqrt{
			\frac{\Lambda_{\mathrm{geom}}}
			{1+\chi(\omega)}
		}
	\end{equation}
}
Having established the analytical foundation of the MBFV framework, culminating in the final expression for the breathing mode frequency (Eq. 36), the framework is now applied across ten distinct viruses embedded in three contrasting media.

\section{Results and Discussion}

The fundamental acoustic vibration of a viral particle, often referred to as the breathing mode or the l= 0, is specifically focused upon because it represents a purely radial motion where the entire viral capsid symmetrically expands and contracts in unison[4-12]. The uniform expansion and contraction serve as a highly accurate indicator of the particles overall structural elasticity. The specific viruses analyzed in this study were selected to represent a diverse range of geometric configurations, spherical, cylindrical, and icosahedral shapes as well as varying physical dimensions. This selection allows for a comprehensive evaluation of how size and structural geometry dictate vibrational frequencies.  
In the field of low frequency Raman scattering (LFRS), there exists an interaction between the external stimuli and viral particles, and they also show mechanical vibration[16-21]. The sound waves confined within the viral capsids cause vibrations. When that frequency of the stimulus aligns with the particle's natural frequency, the viral particle starts to vibrate. This phenomenon is governed by the particle's geometry, radius and density. The inverse relation between the size of a virus and the frequency of a normal mode is observed similarly to the Raman frequency shift calculated from the eigenvalue equation [31-36].
The low frequency breathing mode of ten viruses of different shapes in different mediums was calculated using the formulations developed and discussed in the previous section. Among the viruses are Nipah, Rubella, TMV, Ebola, Rhino, Hepatitis C, HSV-1, SARS-CoV 2, Hepatitis A. The surrounding environments are categorized into three types of media: Water, Glycerol, Biological Habitats.

\begin{table*}[t!]
	\centering
	\caption{Input Parameters: Structural and physical input parameters across various viral geometries.}
	\label{tab:input_parameters}
	\resizebox{\textwidth}{!}{%
		\begin{tabular}{llccccc}
			\toprule
			Geometry & Virus name & Outer capsid & Inner capsid & Capsid density & Genome density & Sound velocity \\
			& & radius (nm) & radius (nm) & ($\mathrm{kg/m^3}$) & ($\mathrm{kg/m^3}$) & ($\mathrm{m/s}$) \\
			\midrule
			Spherical & (1) Nipah & 70 & 55 & 1350 & 1700 & 1800 \\
			Spherical & (2) Rubella & 35 & 25 & 1350 & 1700 & 1800 \\
			Cylindrical & (3) TMV & 8.5 & 2 & 1360 & 1700 & 2200--2400 \\
			Cylindrical & (4) Ebola & 20--25.5 & 8 & 1350 & 1500--1700 & 1800--2000 \\
			Spherical & (5) CMV & 14.4--15.25 & 8.5 & 1360--1370 & 1615--1700 & 1800 \\
			Spherical & (6) Rhino & 15 & 11 & 1350 & 1700 & 1800 \\
			Spherical & (7) Hepatitis C & 30 & 22 & 1350 & 1700 & 1800 \\
			Icosahedral & (8) HSV-1 & 75 & 50--60 & 1350 & 1700 & 2200--2400 \\
			Icosahedral & (9) SARS-CoV2 & 60--80 & 40--50 & 1150--1200 & 1700 & 2200--2400 \\
			Spherical & (10) Hepatitis A & 15 & 11 & 1350 & 1700 & 1800 \\
			\bottomrule
		\end{tabular}%
	}
	
	\vspace{2mm}
	\begin{minipage}{\textwidth}
		\footnotesize
		\textit{*The labels (1)--(10) for each virus correspond to the bar graph in Figure 1.}
	\end{minipage}
\end{table*}

Table 1 elucidates the correlation between some specific parameters and the analysis of viral structures, particularly in the context of Raman spectroscopy and low frequency vibrational modes. The quantities in Table 1 are fundamental structural, mass, and acoustic parameters which are utilized as input for calculations across various viral geometries. The physical scale of each virus is defined by its outer capsid radius and inner capsid radius (nm). The difference between these two values dictates the thickness of the protein shell. This distinction is critical for modelling, as it accounts for the physical space occupied by the genetic material (the core) versus the protective outer layer (the capsid). For example, while Nipah and HSV-1 have large outer radii (70nm and 75nm, respectively), their substantial inner radii (55nm and 50-60 nm) indicate relatively thick capsid walls that significantly influence their overall mechanical stiffness. 

The model incorporates specific mass densities for both the protein shell(ps) and the internal genome core(rho c). A consistent trend across nearly all observed viruses is the higher density of the genetic core (typically $1700 \text{ kg/m}^3$) compared to the protein capsid. The capsid density remains largely uniform, clustering around 1350 to $1370 \text{ kg/m}^3$ for most spherical and cylindrical viruses. Notably, SARS-CoV-2 presents an exception with notably lower capsid density ($1150-1200 \text{ kg/m}^3$), which likely plays a distinct role in its specific low-frequency breathing modes. The acoustic properties of the viral structures, which directly dictate how vibrational waves travel through the capsid, are represented by the longitudinal sound velocity (m/s). Data reveals that sound velocity is highly dependent on viral geometry[45]. Standard spherical viruses (such as Nipah, Rubella and Rhinovirus) generally exhibit a baseline velocity of 1800 m/s. In contrast, rigid icosahedral structures (like CMV, HSV-1 and SARS-Cov-2) and tightly packed cylindrical geometries (like TMV) support faster wave propagation, with sound velocities reaching up to 2200-2400 m/s. Henceforth, these parameters are crucial for effectively calculating the breathing mode frequencies using the proposed framework.

\begin{table}[t]
	\centering
	\caption{The density of different media in which Viral particle is immersed}
	\begin{tabular}{lc}
		\toprule
		Medium & Density ($kg/m^3$) \\
		\midrule
		Phloem sap & 1080 [74] \\
		Water & 1000 [75] \\
		Glycerol & 1260 [75]\\
		Mucus & 1040 [76] \\
		Saliva & 1003 [77] \\
		Blood & 1060 [78] \\
		Cell sap & 1025 [79]\\
		Sewage & 1005 [80] \\
		\bottomrule
	\end{tabular}
\end{table}

It can be observed in Table 2 that different mediums have distinct densities, wherein the viral particles are immersed for Simulated Raman spectroscopic characterizations. This data is fundamental to understand how the surrounding media (environment) influences the vibrational spectroscopic of the viral particles. Ranging from pure water ($1000 \text{ kg/m}^3$) to physiologically relevant biological fluid (blood, saliva, mucus) and environmental conditions (phloem sap, sewage). The density variation ($1000-1260 \text{ kg/m}^3$) across different media creates acoustic and mechanical environments that directly affect Raman spectroscopic magnitudes. The Raman spectroscopy should work reasonably well in biological fluids because the density difference however small can significantly influence its resonant properties[10,69].
\begin{table*}[t]
	\centering
	\caption{The calculated breathing mode frequencies for the ten viruses of this study using MBFV, immersed in different media, compared against available experimental and theoretical literature values. Four additional viruses (HBV, M13, BMV, CCMV) are included for validation while their geometries are identified from the literature.}
	\footnotesize
	\setlength{\tabcolsep}{3.5pt}
	\begin{tabular}{
			p{1.5cm}
			p{1.6cm}
			p{1.4cm}
			ccc
			p{2.1cm}
			p{2.1cm}
			p{1.6cm}
		}
		\toprule
		Geometry & Virus & Habitat & \multicolumn{3}{c}{Breathing mode frequency (GHz)} & \multicolumn{3}{c}{Comparative Analysis (GHz)} \\
		\cmidrule(lr){4-6} \cmidrule(lr){7-9}
		& & & Habitat & Glycerol & Water & Experimental & Previous Model & Calculated (MBFV) \\
		\midrule
		Spherical & Nipah & Saliva & 14.85 & 14.32 & 14.68 & - & - & - \\
		Spherical & Rubella & Mucus & 30.35 & 29.38 & 30.54 & - & - & - \\
		Cylindrical & TMV & Cell sap & 60.475 & 58.58 & 60.05 & 60 [24] & 63 [24] & 60.023 \\
		Cylindrical & Ebola & Blood & 30.04 & 28.95 & 30.40 & - & - & - \\
		Spherical & CMV & Phloem sap & 85.80 & 83.59 & 87.56 & - & - & - \\
		Spherical & Rhino & Nasal mucus & 70.32 & 68.09 & 70.75 & - & - & - \\
		Spherical & Hepatitis C & Blood & 35.05 & 34.04 & 35.37 & - & - & - \\
		Icosahedral & HSV-1 & Saliva & 18.07 & 17.40 & 18.08 & - & - & - \\
		Icosahedral & SARS-CoV-2 & Mucus & 19.12 & 18.43 & 19.25 & - & - & - \\
		Spherical & Hepatitis A & Sewage & 70.69 & 68.09 & 70.75 & - & - & - \\
		\midrule
		Icosahedral & HBV & - & - & - & - & 25.78-190.67 [9] & - & 53.20 \\
		Cylindrical & M13 & - & - & - & - & 254 [36] & $<$599.58 [36] & 251 \\
		Spherical & BMV & - & - & - & - & 58.5 [23] & - & 58.33 \\
		Spherical & CCMV & - & - & - & - & 35.67-509.65 [9] & - & 73.67 \\
		\bottomrule
	\end{tabular}
\end{table*}

\begin{figure}[t]
	\centering
	\includegraphics[width=\columnwidth]{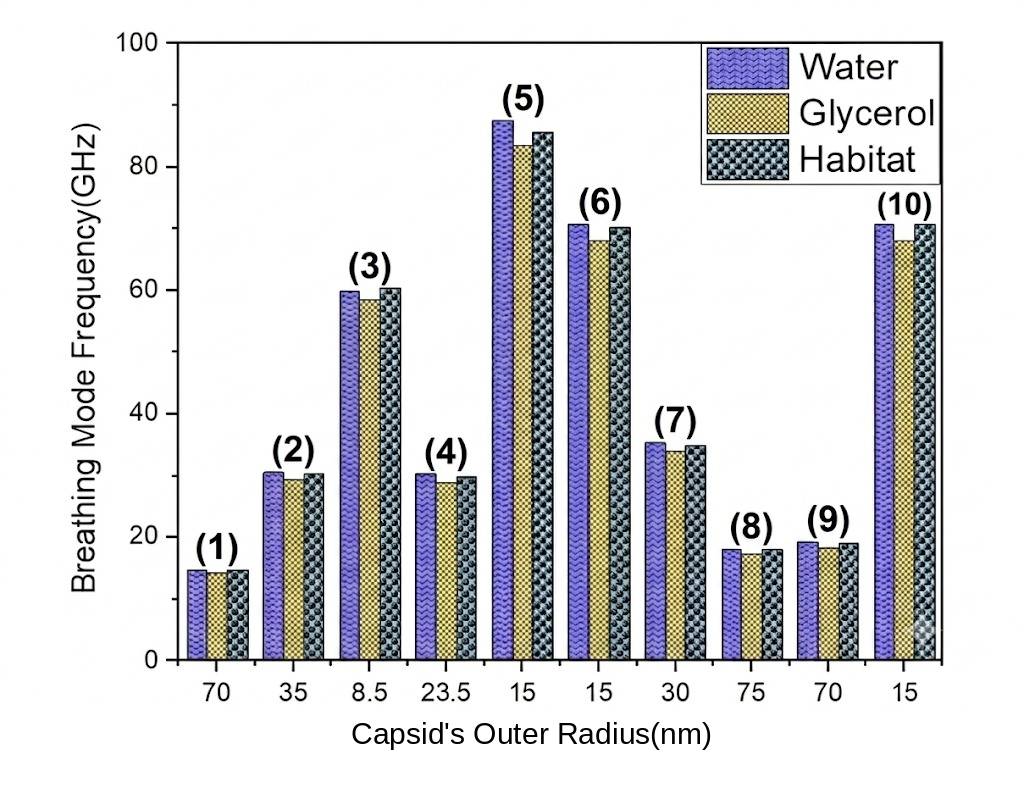} 
	\caption{Clustered bar chart illustrating a comparative analysis across these media}
\end{figure}

The data, illustrated in Fig 1 and detailed in Table 3, highlight a comparative analysis in three distinct media, water (a standard low viscosity baseline medium), glycerol (a high-viscosity damping medium), and the specific natural biological habitat of each virus. The most pronounced trend observed in our data is the strict inverse relationship between a virus's outer radius and its breathing mode frequency. As we see, that virus with a 15nm radius, such as the CMV, Rhino and Hepatitis A, exhibits the highest frequency ranges in the study, peaking between 68.09 and 87.56 GHz depending on the medium. In contrast, larger viruses, such as Nipah and SARS-COV-2, which correspond to 70nm clusters, demonstrate much lower breathing mode frequencies, 14.32 GHz and 19.25 GHz range. Table 3 represents the Breathing mode frequencies demonstrated by the viral shells when they are embedded in the contrasting liquid medium. The breathing mode displays the characteristic resonance vibrations that occur within the viral particle structures. 

As seen in Table 3, when the viral particle is immersed in distinct mediums, we get different frequencies. This can be observed with regards to the calculated frequencies, which strongly indicate that none of the viral shells has fixed rate of vibrations. The surrounding media plays a crucial role by affecting the viral particles as their density and viscosity effectively dampen the vibrations of the aforementioned viral particles. The third medium in the table is incorporated to facilitate the behavioural dynamics of the viral particles embedded within their native biological habitat. Upon observing the data, the wide range across the different viruses(around 14.32 to 87.56 GHz) is precisely noticeable. These viruses have idiosyncratic structural properties (rigidity and elasticity) which results in different capsid sizes, protein compositions, and unique geometries that produce different resonant peaks. It can be clearly observed that  CMV has the highest frequency (approx. 87 GHz) and the Nipah virus has the lowest frequency range (approx. 14 GHz). This trend is due to the fact that each viral particle has unique prominent structural characteristics representing a bar graph of breathing mode frequency vs the outer capsid radius of the virus. We consider all the three mediums previously utilized for the same (Water, Glycerol and Habitat). In this we can perceive that Nipah has an outer capsid radius of approximately 70 nm and its breathing mode frequency is roughly around 14 GHz, whereas CMVs outer capsid radius is nearly 15 nm while having a breathing mode frequency of 87 GHz. The relation which can be inferred is that; larger the viral capsid, the slower the oscillations and vice versa. The Comparative Analysis of Table 3 situates the MBFV calculations against the experimental and previous theoretical models. The calculated (MBFV) values reported here were evaluated using the same medium density reported in each corresponding reference. For TMV, the experimental value, 60 GHz, was measured by stimulated low-frequency Raman scattering (SLFRS) in a Tris-HCl buffer suspension [24]. The M13 bacteriophage data were obtained by Raman spectroscopy in water as the medium [36]. The BMV measurement of 58.5 GHz was observed in a phosphate buffer medium [23]. Finally, the HBV and CCMV , both sourced from atomistic normal-mode calculations [9]. Therefore, this variation of breathing mode frequencies with medium depicts that the breathing-mode frequency of a vius cannot be treated as a fixed, medium-independent property. Furthermore, we present the calculated low-frequency Raman spectra for these viruses when embedded in different media(in figures 2,4,6). These media are divided into three specific categories, designated Set 1, Set 2, and Set 3.

\section*{Set 1: Water Medium}

Figure 2 shows the variation in the peak intensity that suggested differences in the efficiency of Raman scattering and the interaction strength among the viruses. Raman spectroscopy avoids significant solvent interference, yielding clearer and more reliable vibrational spectra[29,32,34,35,38]. The line width of the spectral line indicated variations in damping behaviour and resonance stability[11, 46, 23, 54, 71]. Narrow spectral features imply highly localized and well defined vibrational states[9, 40, 51, 19]. The graph primarily establishes a spectral database capable of distinguishing viral species based on resonant characteristics. The graph compares the Raman response of different viruses under identical environmental conditions precisely, to monitor whether each virus exhibits a unique resonant frequency or not and consequently, how different breathing modes interact. The result shows that the sharpest peak is for SARS-CoV, which has maximum intensity, whereas CMV has the lowest intensity. Henceforth, it can be stated that low frequency vibrational modes are indicative of collective acoustic phonons, like oscillations across extended viral structures. In general, larger viral particles exhibit reduced elastic stiffness, leading to lower frequency vibrational responses and, for high frequency vibrational modes, this corresponds to shorter wavelengths in the localized vibrational mode [10,41]. These are well associated with structurally rigid, compact viral particles, or rather, they have very small dimensional viral properties. Figure 3 represents the normalized Raman spectra of the viral breathing mode. During  normalization, all spectral peaks are scaled relative to their maximum intensity, this approach facilitates an accurate comparison of resonant behaviour between different viral species. The normalized spectra reveal variations in spectral broadening and sharpness in resonant peaks. Narrow spectral peaks indicate high resonant behaviour associated with larger quality factors (Q factor),which reflect lower energy dissipation,  whereas broad peaks indicate stronger damping effects, ultimately causing an increase in energy loss and reduced resonance selectivity [10]. From this information, the parameters we have extracted from the graph are; the peak position, peak width and spectral broadening, relative damping behaviour and quality factor characteristics.

\begin{figure}[t]
	\centering
	\includegraphics[width=\columnwidth]{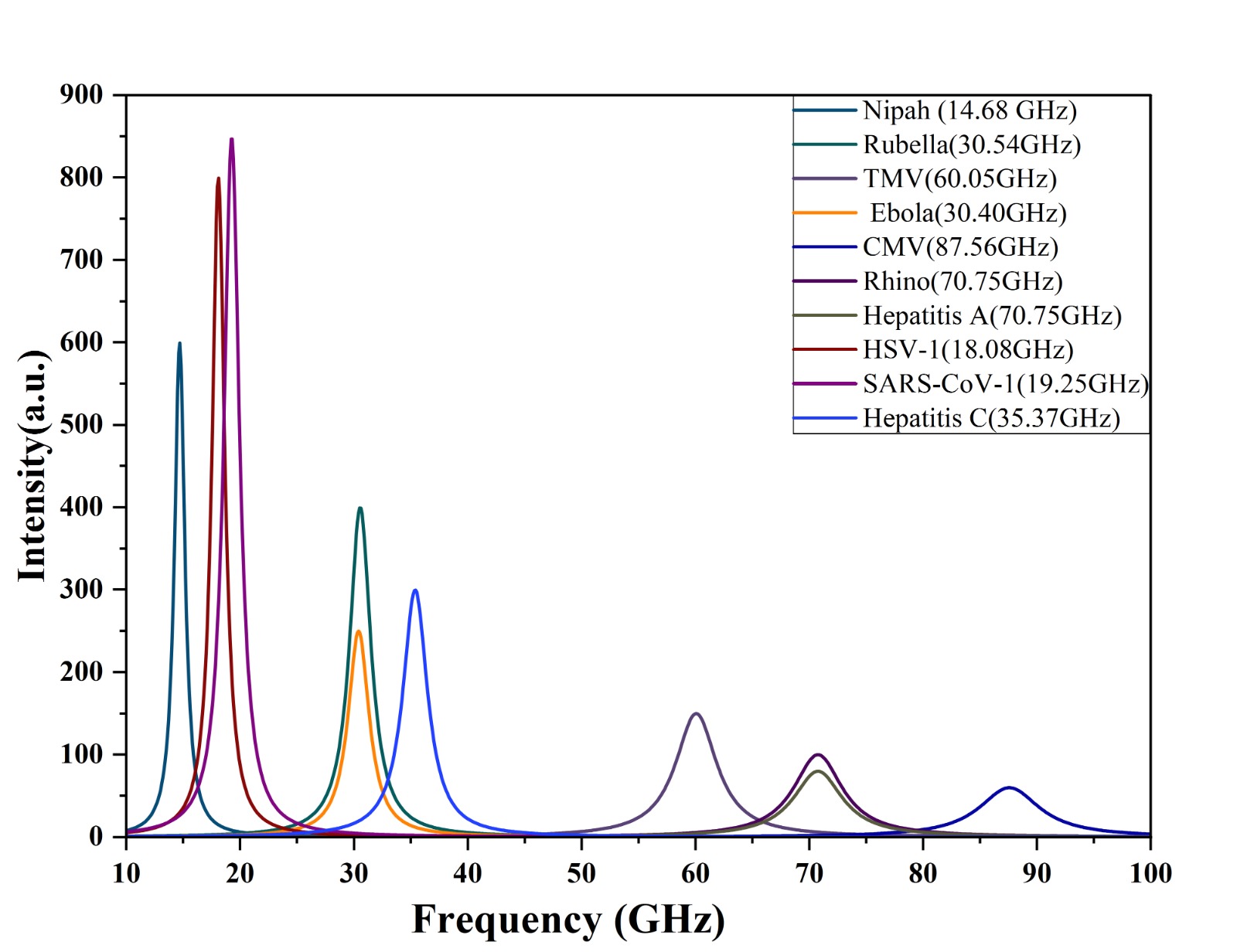} 
	\caption{Viral Breathing Mode Raman Spectra (Water)}
\end{figure}

\begin{figure}[t]
	\centering
	\includegraphics[width = 10cm]{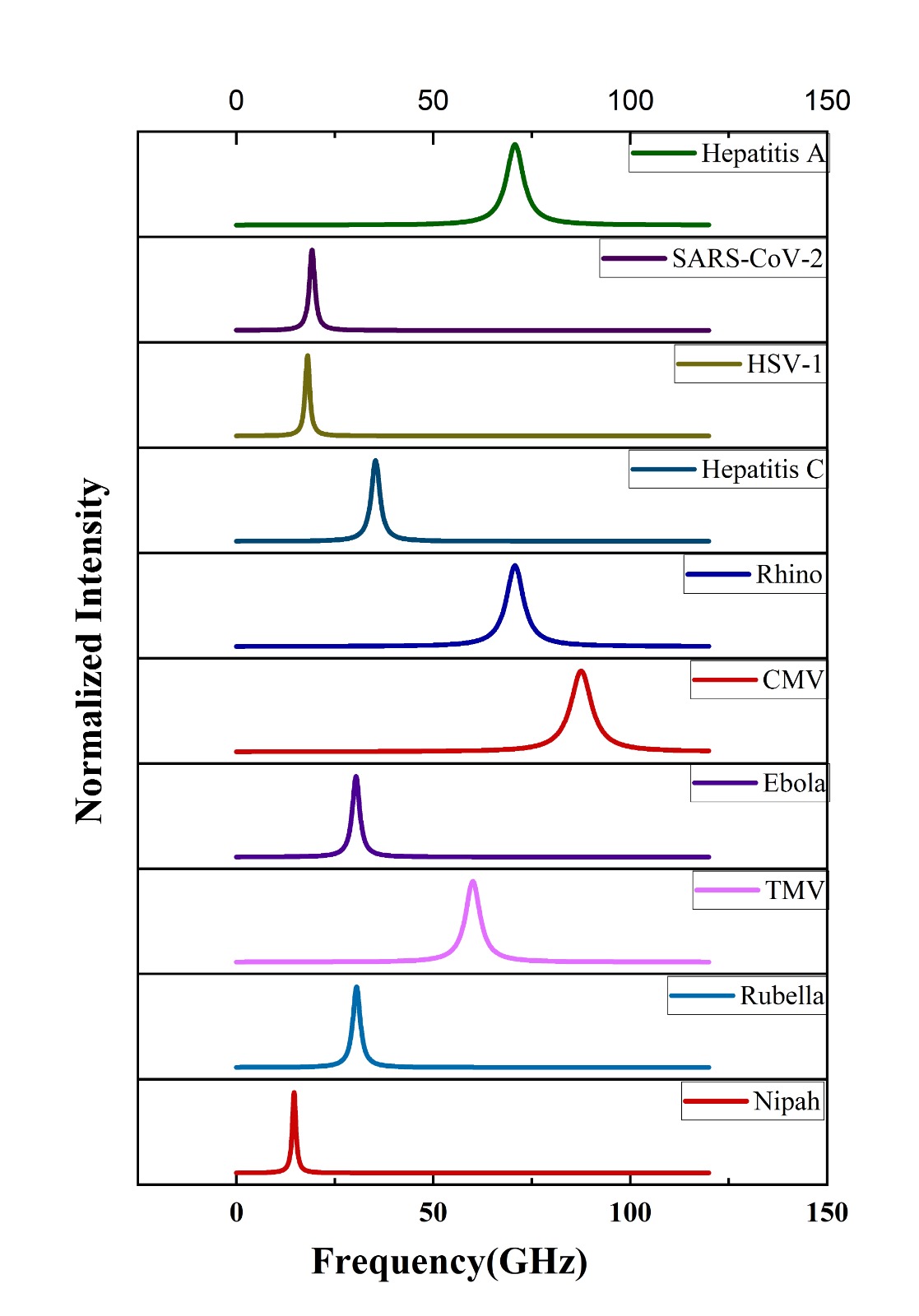}
	\caption{Normalized Viral Breathing Mode Raman Spectra (Water)}
\end{figure}

\section*{Set 2: Glycerol Medium}

Glycerol is generally regarded as one of the most effective biological cryopreservants. For the medium, we observe distinct resonant peaks for all evaluated viral particles as seen in Figure 4. The resonant frequencies span approximately 14.32 to 83.59 GHz, indicating a variation in structural and mechanical properties among the viral models. The lower frequency region contains Nipah (14.32 GHz), HSV-1 (17.40 GHz), and SARS-CoV-2 (18.43 GHz), which are indicative of collective long wavelengths and vibrational oscillations associated with large scale structures and comparatively lower effective stiffness. These viruses exhibit relatively low resonant frequencies because their vibrational behaviour is governed by weaker restoring forces. High frequency peaks are observed with some viral particles like SAR-CoV-2 and HSV 1. The increase in resonant frequency indicates stronger elastic restoring force, greater structural rigidity, and more confined vibrational behaviour. The influence of the glycerol medium is particularly important because its higher viscosity compared to water introduces very strong damping[10]. The increased viscosity modifies the vibrational response by affecting resonance broadening and energy dissipation. Peaks with broader linewidths indicate increased damping and reduced vibrational stability, whereas narrow peaks correspond to highly localized resonant behaviours.

Figure 5 represents the normalized intensity graph of Figure 4, in which all spectral peaks are scaled relative to their respective maximum intensities in order to compare the resonant behaviour. The normalization depicts that while the breathing mode frequencies remain consistent with those reported in Figure 4, the spectral widths vary considerably across the viruses, showing differences in their damping characteristics in the glycerol medium. Viruses such as SARS-CoV-2 and HSV-1 have comparatively narrow peaks even after normalization, indicating higher Q-factors. In contrast, larger capsids such as Nipah and Ebola exhibit broader spectra.

\begin{figure}[t]
	\centering
	\includegraphics[width=\columnwidth]{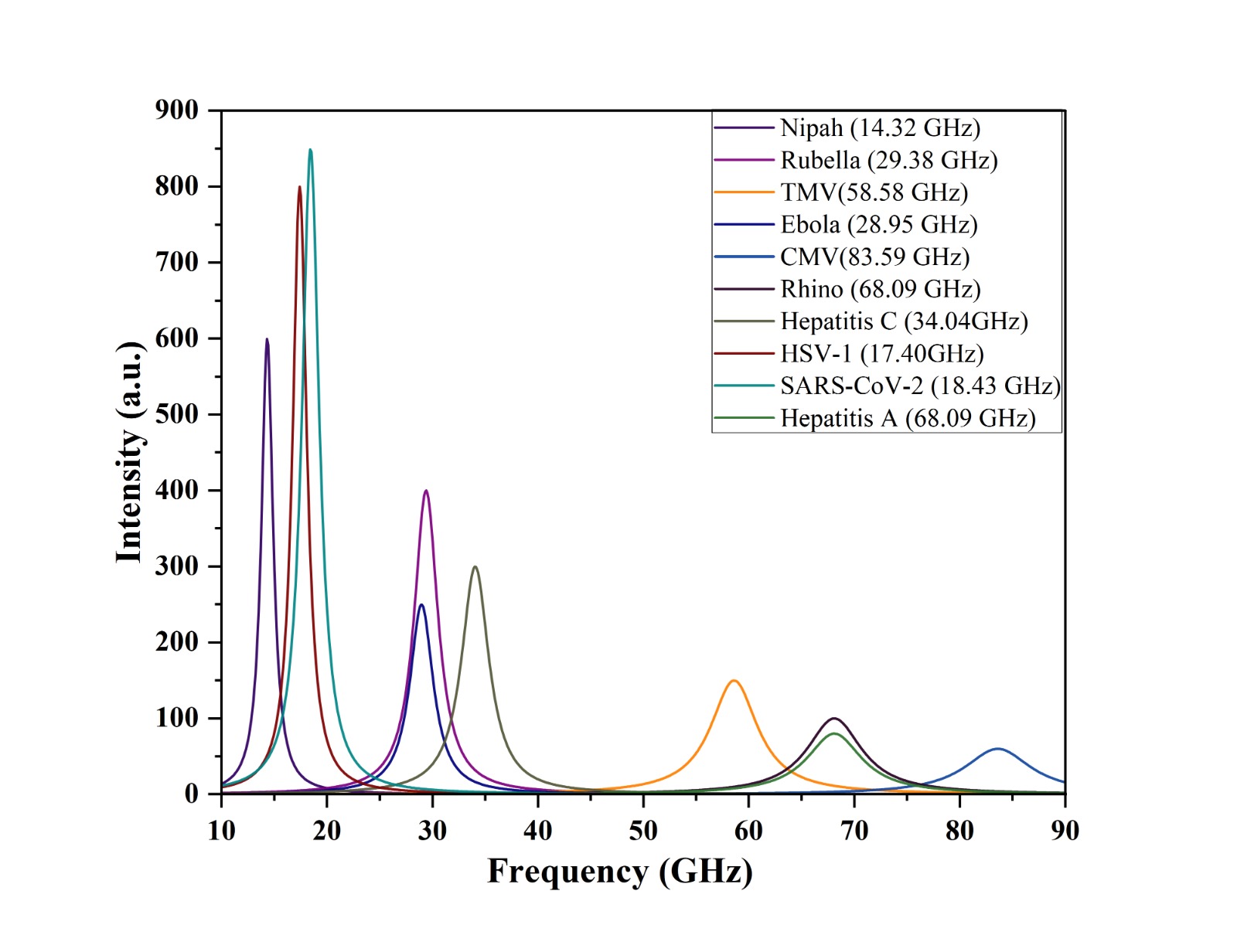} 
	\caption{Viral Breathing Mode Raman Spectra (True Glycerol)}
\end{figure}

\begin{figure}[t]
	\centering
	\includegraphics[width = 10cm]{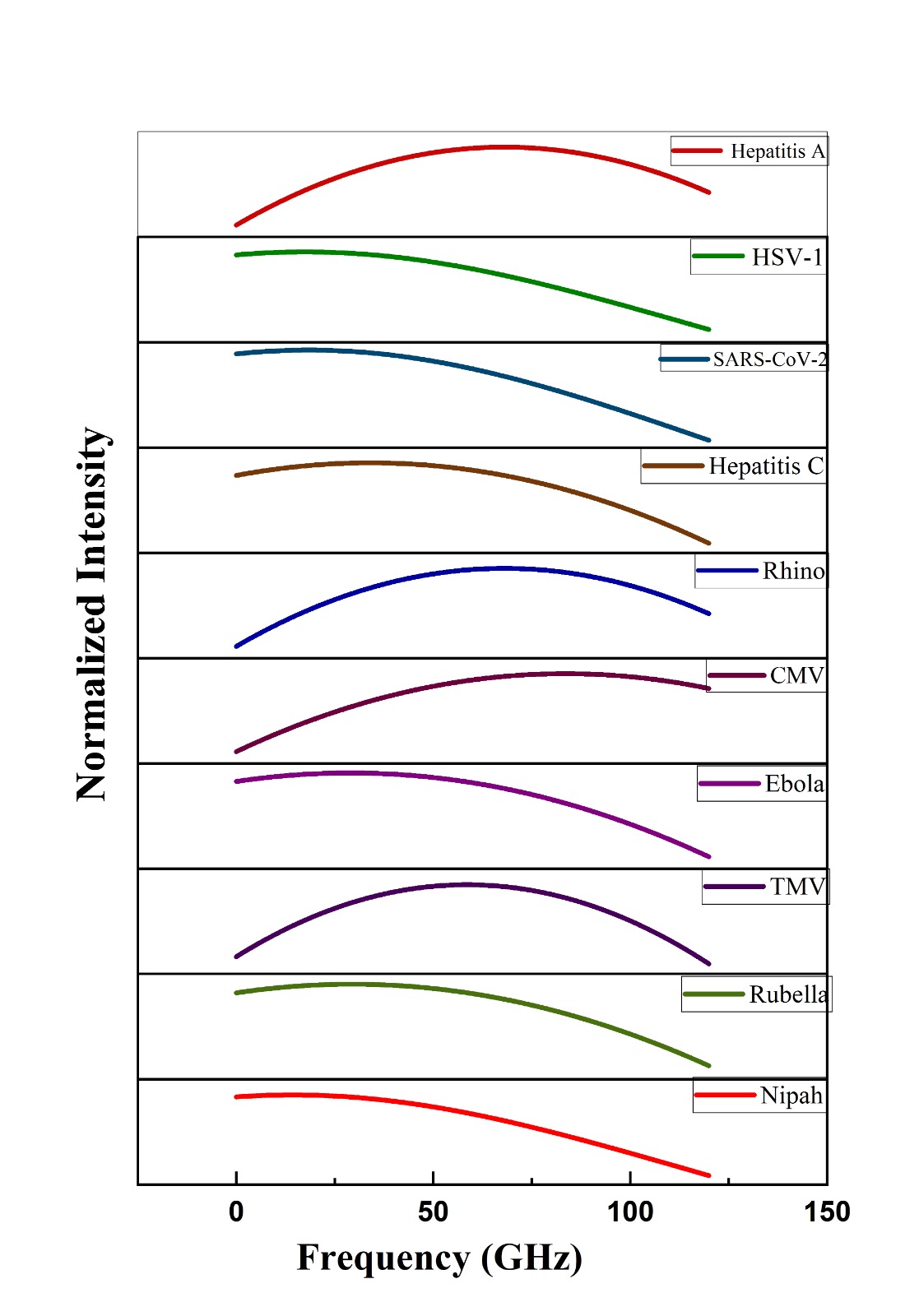}
	\caption{Normalized Viral Breathing Mode Raman Spectra (True Glycerol)}
\end{figure}

\section*{Set 3: Biological Habitat}

The Raman spectra shown in this Figure 6 were created to explore how different biological environments affect the vibrational behaviour of viral breathing modes while considering the realistic structural properties of the viruses. Unlike previous studies that were confined to a common medium, this model includes realistic environments such as blood, mucus, and saliva, providing a biological representation of viral dynamics. Further it displays a composite representation of Raman spectroscopy for ten viral particles. Each spectral peak represents the characteristic breathing mode's response of an individual viral particle in its respective milieu, with the peak position directly related to the inherent size of the virion. The sharpness of the peaks vary among viruses. SARS-CoV-2 and HSV-1 exhibit sharp peaks and high Q-factors, while Hepatitis A and CMV exhibit broader peak and low Q-factors, indicating stronger damping effects. The normalized Raman spectroscopic analysis of viral breathing modes in biological habitats. through normalization , all spectral peaks have maximum intensity, thus this can easily be interpreted in spectral broadening and resonant behaviour patterns.

Figure 7 represents the normalized Raman spectra of all ten viruses in their respective biological habitats, in which each spectral peak is scaled to its maximum intensity. The normalized spectra shows a heterogeneity in the peaks of the viruses due to the contrasting viscoelastic properties of their individual biological media. Viruses residing in comparatively dilute fluids such as saliva exhibit narrower widths and higher Q-factors, indicative of reduced acoustic energy dissipation, whereas those in macromolecule rich environments such as mucus and blood display appreciably broadened spectra due to the elevated viscous drag.

\begin{figure}[t]
	\centering
	\includegraphics[width=\columnwidth]{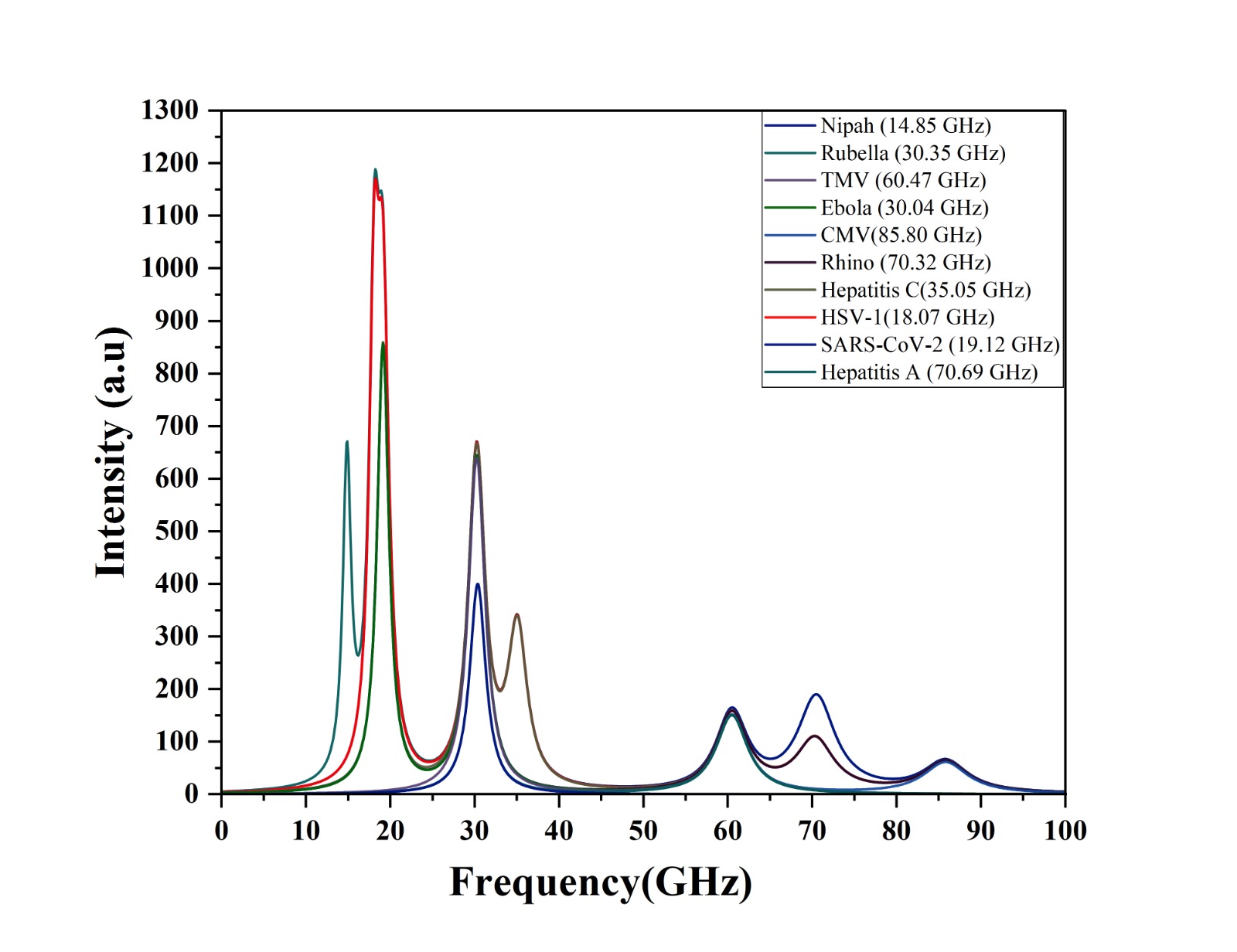} 
	\caption{Viral Breathing Mode Raman Spectra (Biological Habitat)}
\end{figure}

\begin{figure}[t]
	\centering
	\includegraphics[width = 10cm]{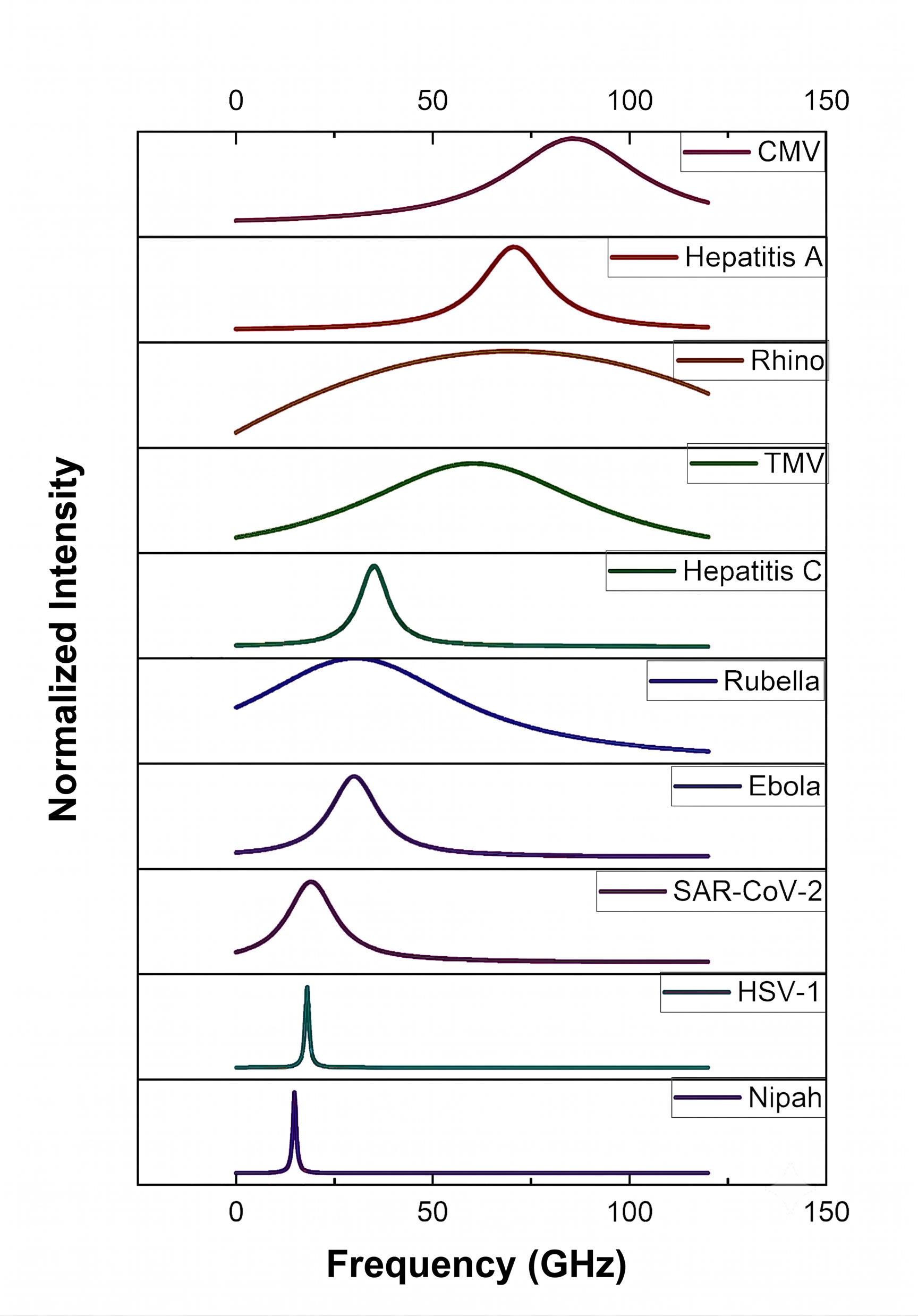}
	\caption{Normalized Viral Breathing Mode Raman Spectra (Biological Habitat)}
\end{figure}

\begin{itemize} 
	\item The spectral variations observed across the different media are governed by the acoustic radiation damping of the viral capsid acting as a nanomechanical oscillator within a continuous viscoelastic fluid.The spectral bandwidth is defined as the Full Width at Half Maximum (FWHM, or $w$) which is inversely proportional to its Quality Factor ($Q$), dictated by:
	$$w = \frac{f_c}{Q}$$
	Here, $f_c$ is the frequency of the breathing mode . The $Q$ factor itself is highly dependent on the density ($\rho_m$) and the dynamic viscosity of the surrounding media ($\eta$).
	The normalized intensity figures 3,5,7 can be  articulated respectively:
	In Water (Underdamped): water possesses low density ($\rho \approx 1000$ kg/m$^{3}$) and low dynamic viscosity, which minimizes the acoustic energy dissipation into the surrounding medium. Consequently, the viral oscillator maintains a high $Q$ factor, resulting in slender spectral bandwidths (sharp Lorentzian peaks).
	In Glycerol (Overdamped): Glycerol exhibits higher density ($\rho \approx 1260$ kg/m$^{3}$) and extreme dynamic viscosity, therefore, the capsid experiences severe viscous, drag and high acoustic radiation coupling. This massive energy dissipation drastically attenuates the $Q$ factor ($Q \ll 1$). 
	In Biological Habitats (Transitional/Mixed): Bio-fluids represent an intermediate viscoelastic state. Dilute fluids such as saliva yield lower acoustic damping, which results in sharper peaks, whereas macromolecule-rich fluids such as mucus and blood induce moderate viscous drag, yielding proportionally broadened FWHM profiles.
	It should be thoroughly noted that, normalization $\left(\frac{I}{I_{max}} = 100\%\right)$ does not physically alter the FWHM, as it is a mathematical scalar.
	
	Table 4 represents MBFV-calculated breathing-mode frequencies against available experimental and prior models across a range of radii and geometries of viruses. The structural parameters like radius, capsid density, genome density and longitudinal sound velocity used for each MBFV calculation are sourced from the corresponding experimental reference for that specific virus, making it comparable. It is to be noted that there are many viruses for which there is no prior data available for breathing mode frequency [10,12,16,22,24,34,67]. Finally, this table also demonstrates the accuracy of MBFV calculations with respect to experimental (LFRS/SLFRS) values.

\end{itemize}

\renewcommand{\arraystretch}{1.3}
\begin{table*}[t]
	\centering
	\caption{%
		Cross-literature comparison of breathing-mode ($l{=}0$) frequencies
		organised by increasing outer capsid radius. MBFV values are computed
		in this work for water and glycerol media using virus-specific
		structural parameters sourced from literature.
	}
	\footnotesize
	\setlength{\tabcolsep}{3pt}
	\begin{tabular}{
			p{2.0cm}
			p{1.5cm}
			p{1.3cm}
			p{1.1cm}
			p{1.2cm}
			p{2.3cm}
			p{1.1cm}
			p{2.5cm}
			p{1.2cm}
		}
		\toprule
		\multirow{2}{*}{\shortstack[l]{Species}}
		& \multirow{2}{*}{Geometry}
		& \multirow{2}{*}{\shortstack[l]{Radius\\$R$ (nm)}}
		& \multicolumn{2}{c}{MBFV (GHz)}
		& \multicolumn{2}{c}{Exp.\ (GHz)}
		& \multicolumn{2}{c}{Prev.\ Models (GHz)} \\
		\cmidrule(lr){4-5}\cmidrule(lr){6-7}\cmidrule(lr){8-9}
		& & & Water & Glycerol & Water/buffer & Glycerol & Water & Glycerol \\
		\midrule
		
		Generic
		& Spherical & ${\sim}5.6$
		& 184.051 & 173.048
		& ---
		& ---
		& ${\sim}$140 \cite{Krishnam2017}
		& --- \\[2pt]
		
		TMV
		& Cylindrical & 8.5
		& 60.34 & 57.38
		& 60 \cite{Donchenko2017}
		& ---
		& 63 \cite{Donchenko2017}; 62.97 \cite{Balandin2005}
		& --- \\[2pt]
		
		BMV
		& Spherical & ${\sim}14$
		& 58.89 & 56.93
		& 58.5 \cite{Sirotkin2010}
		& ---
		& ---
		& --- \\[2pt]
		
		CMV, CCMV, Poliovirus
		& Spherical & ${\sim}14$--15
		& 87.59 & 83.96
		& ---
		& ---
		& 35.67--509.65 \cite{Dykeman2010a}
		& --- \\[2pt]
		
		HBV
		& Icosahedral & ${\sim}16$
		& 66.69 & 64.16
		& 25.78--190.67 \cite{Dykeman2010a}
		& ---
		& ---
		& --- \\[2pt]
		
		CaMV
		& Icosahedral & ${\sim}17.5$
		& 60.42 & 58.16
		& 58.2 \cite{Sirotkin2014}
		& ---
		& ---
		& --- \\[2pt]
		
		Ebola
		& Cylindrical & $20$--$25.5$
		& 30.40 & 28.95
		& ---
		& ---
		& ---
		& --- \\[2pt]
		
		Hepatitis C
		& Spherical & 30
		& 35.37 & 34.04
		& ---
		& ---
		& ---
		& --- \\[2pt]
		
		Rubella
		& Spherical & 35
		& 30.54 & 29.38
		& ---
		& ---
		& ---
		& --- \\[2pt]
		
		Generic
		& Spherical & ${\sim}50$
		& 24.65 & 23.32
		& ---
		& ---
		& ${\sim}$17 \cite{Talati2006}
		& ${\sim}$16.5 \cite{Talati2006} \\[2pt]
		
		Generic
		& Spherical & ${\sim}56$
		& 21.90 & 20.72
		& ---
		& ---
		& ${\sim}$14 \cite{Krishnam2017}
		& --- \\[2pt]
		
		SARS-CoV-2, LentiGFP
		& Icosahedral & $60$--$80$
		& 19.25 & 18.43
		& 19--22 \cite{Zhang2025b}
		& ---
		& ---
		& --- \\[2pt]
		
		Nipah
		& Spherical & 70
		& 14.68 & 14.32
		& ---
		& ---
		& ---
		& --- \\[2pt]
		
		HSV-1
		& Icosahedral & 75
		& 18.08 & 17.40
		& ---
		& ---
		& ---
		& --- \\
		
		\bottomrule
	\end{tabular}
\end{table*}
\clearpage
\section{Conclusion}

The present paper reports a model beyond the three generations of approaches to investigate the low frequency breathing modes of viruses of different shapes, sizes, and surrounding media. The proposed framework segregates viral particles into appropriate coordinate systems according to their intrinsic morphology and corresponding geometry-specific eigenvalues. The model predicts breathing mode frequencies across a range of 14.32 to 87.56 GHz without requiring atomistic parameters or mesh-dependent boundary conditions. Validation against experimental LFRS data for TMV, M13, and BMV confirms close quantitative agreement, with the MBFV consistently performing more accurately than the pre-existing theoretical approaches, which have systematic overestimation due to the omission of solvent damping effects. The framework further captures the medium-dependent shift in spectral lines, producing the transition from underdamped sharp resonances in water to overdamped broadened profiles in glycerol and biological fluids. These results establish that MBFV is computationally efficient and a geometrically robust tool for the mechanical fingerprinting of viral particles in diverse biological environments. The proposed Multigeometric Breathing mode Framework for Virus (MBFV), yields a mean absolute relative error of  0.503\%. Ultimately, MBFV serves as a highly accurate tool for aiding in the developmental methods for
viral identification and potential elimination of these particles by systematically targeting their breathing modes.

\section*{Conflicts of interest}
There are no conflicts to declare.

\section*{Acknowledgement}
This research was carried out with the support of the DST FIST(SR/FST/PS-I/2022/230) for the computational facilities developed grant and the Anusandhan National Research Foundation (ANRF-SERB), Government of India, for financial support.



\end{document}